\documentstyle[12pt,aasms4,psfig]{article}
\newcommand{\etal}{{\it et al.\ }}
\newcommand{\grados}{^\circ}

\lefthead{Carrera \etal}
\righthead{Ursa Minor}

\begin{document}

\title{The Star Formation History and the spatial distribution of stellar
populations in the Ursa Minor Dwarf Spheroidal Galaxy}

\author{Ricardo Carrera} 
\affil{Instituto de Astrof\'\i sica de Canarias,
E38205-La Laguna, Tenerife, Canary Islands, Spain}

\author{Antonio Aparicio} 
\affil{Instituto de Astrof\'\i sica de Canarias,
E38205-La Laguna, Tenerife, Canary Islands, Spain}
\affil{Departamento de Astrof\'\i sica, Universidad de La Laguna,
E38200 - La Laguna, Tenerife, Canary Islands, Spain}

\author{David Mart\'\i nez-Delgado} 
\affil{Instituto de Astrof\'\i sica de Canarias,
E38205-La Laguna, Tenerife, Canary Islands, Spain}

\author{Javier Alonso-Garc\'{\i}a} 
\affil{Department of Astronomy, University of Michigan, Ann Arbor, MI 48109}

\begin{abstract}

As a part of a project devoted to the study of the Ursa Minor dSph, the star
formation history of the galaxy is presented in this paper. The analysis uses
wide field photometry, encompassing about $1\grados\times1\grados$ (the total
covered area being $0.75$ deg$^2$), which samples the galaxy out to its tidal
radius. Derivation of the SFH has been performed using the synthetic partial
model technique.

The resulting SFH shows that Ursa Minor hosts a predominantly old stellar
population, with virtually all the stars formed earlier than 10 Gyr ago and 90\% of
them formed earlier than 13 Gyr ago. Nevertheless, Ursa Minor color-magnitude
diagram shows several stars above the main, old turn-off forming a blue-plume
(BP). If these stars were genuine, main-sequence stars, Ursa Minor would have
maintained a low star formation rate extending up to $\sim2$ Gyr
ago. However, several indications (relative amount and spatial distribution
of BP stars and difficulty to retain processed gas) play against this
possibility. In such context, the most reliable hypothesis is that BP stars
are blue-stragglers originating in the old population, Ursa Minor hence
remaining the only Milky Way dSph satellite to host a pure old stellar
population. A marginally significant age gradient is detected, in the sense
that stars in outer regions are slightly younger, in average.

The distance of Ursa Minor, has been calculated using the magnitude of the
horizontal-branch and a calibration based on Hipparcos data of main sequence
sub-dwarfs. We estimated a distance $d=76\pm4$ Kpc, which is slightly larger
than previous estimates. From the RGB color, we estimate a metallicity
$[Fe/H]=-1.9\pm0.2$, in agreement with a previous spectroscopic
determination. No metallicity gradients have been detected across the galaxy.

\end{abstract}


\section{Introduction}

Dwarf spheroidal (dSph) galaxies are objects of high cosmological
significance. In generic hierarchical clustering scenarios for galaxy formation,
such as cold dark matter dominated cosmologies (White \& Rees 1978;
Blumenthal \etal 1984; Dekel \& Silk 1986), dwarf galaxies should have formed
prior to the epoch of giant galaxy formation and would be the building blocks
of larger galaxies. The dSphs observed today would be surviving objects that
have not merged within larger galaxies. Therefore, unveiling their underlying
structure could provide important clues about process of dwarf galaxy
formation that are now observed at high red-shifts. Local Group galaxies offer
the only opportunity of studying their evolution in great detail through the
observation of their resolved stellar populations, which are fossil records
of the star formation history (SFH).

DSph companions of the Milky Way were historically considered as old systems
inhabited by globular cluster-like, population II stars (Baade
1963). However, later works revealed traces of intermediate-age stars, such
as carbon stars (Aaronson \& Mould 1980; Mould \etal 1982; Frogel \etal 1982;
Azzopardi, Lequeux \& Westerlund 1986) and bright AGB stars (Elston \& Silva
1992; Freedman 1992; Lee, Freedman \& Madore 1993; Davidge 1994; but see also
Mart\'{\i}nez-Delgado \& Aparicio 1997). More recent studies, based on the
analysis of color-magnitude diagrams (CMDs) using synthetic CMDs, have shown
that dSph are objects with complex and varied SFHs (Mighell 1997; Mart\'\i
nez-Delgado, Gallart, \& Aparicio 1999; Gallart \etal 1999a; Hern\'andez,
Gilmore \& Valls-Gabaud 2000; Aparicio, Carrera \& Mart\'{\i}nez-Delgado
2001) and the idea that no two dSphs could be identified to have similar SFHs
has become popular.

As part of a bigger project devoted to the study of the Ursa Minor dSph
galaxy (see Mart\'\i nez-Delgado et al. 2001, 2002), we present in this paper
the results on the SFH of this galaxy. Our data sample the galaxy out to
$40'$ from its center, which is of the order of its tidal radius (see Irwin
\& Hatzidimitriou 1995; Kleyna \etal 1998; Mart\'\i nez-Delgado \etal
2001). Ursa Minor is the second closest dwarf spheroidal galaxy
satellite of the Milky Way ($d=69$ Kpc) and was discovered in the Palomar
survey by Wilson (1955). Previous works show an old system similar, in age
and abundance, to the ancient metal-poor Galactic globular cluster M92
(Olszewski \& Aaronson 1985; Mighell \& Burke 1999) and owning a well
populated, predominantly blue horizontal-branch (HB) (see e.g. Cudworth,
Olszewski \& Schommer 1986; Kleyna \etal 1998). Mart\'\i nez-Delgado \&
Aparicio (1998) pointed out the possibility that a marginal intermediate-age
stellar population could exist in the galaxy, associated to the blue-plume
present in its CMD. Olszewski \& Aaronson (1985) and Mighell \& Burke (1999)
derived distance moduli of $(m-M)_0=19.0\pm0.1$ and $19.18\pm0.12$
respectively, from a sliding fit to the M92 ridge line. As for the
metallicity, it has been recently estimated by Shetrone, C\^ot\'e, \& Sargent
(2001), from spectroscopic measurements of giant stars, to be
$[Fe/H]=-1.9\pm0.11$.

Mart\'{\i}nez-Delgado \etal (2001) have shown evidences that Ursa Minor is
in a process of tidal disruption. This motivates the spatially extended study
of the SFH that we present here. Before us, Hern\'andez \etal (2000) (see
also Valls-Gabaud, Hern\'andez, \& Gilmore 2000) have obtained the SFH for
the very central part of Ursa Minor using deep, but small field ($\sim
2\farcm5\times 2\farcm5$) HST data. They find a stellar population mainly
composed of stars older than 10 Gyr.

The paper is organized as follows. In \S 2, the observations and photometric
reduction are presented. In \S 3, the CMD of Ursa Minor is discussed,
followed by the estimation of the distance and metallicity. The rest of the
paper is mainly devoted to the derivation of the SFH and the discussion of
possible evolution scenarios for Ursa Minor, to which \S 4, 5 and 6 are
devoted. Finally, the main results of the paper are summarized in \S 7.

\section{Observations and data reduction}

\begin{figure}
\centerline{\psfig{figure=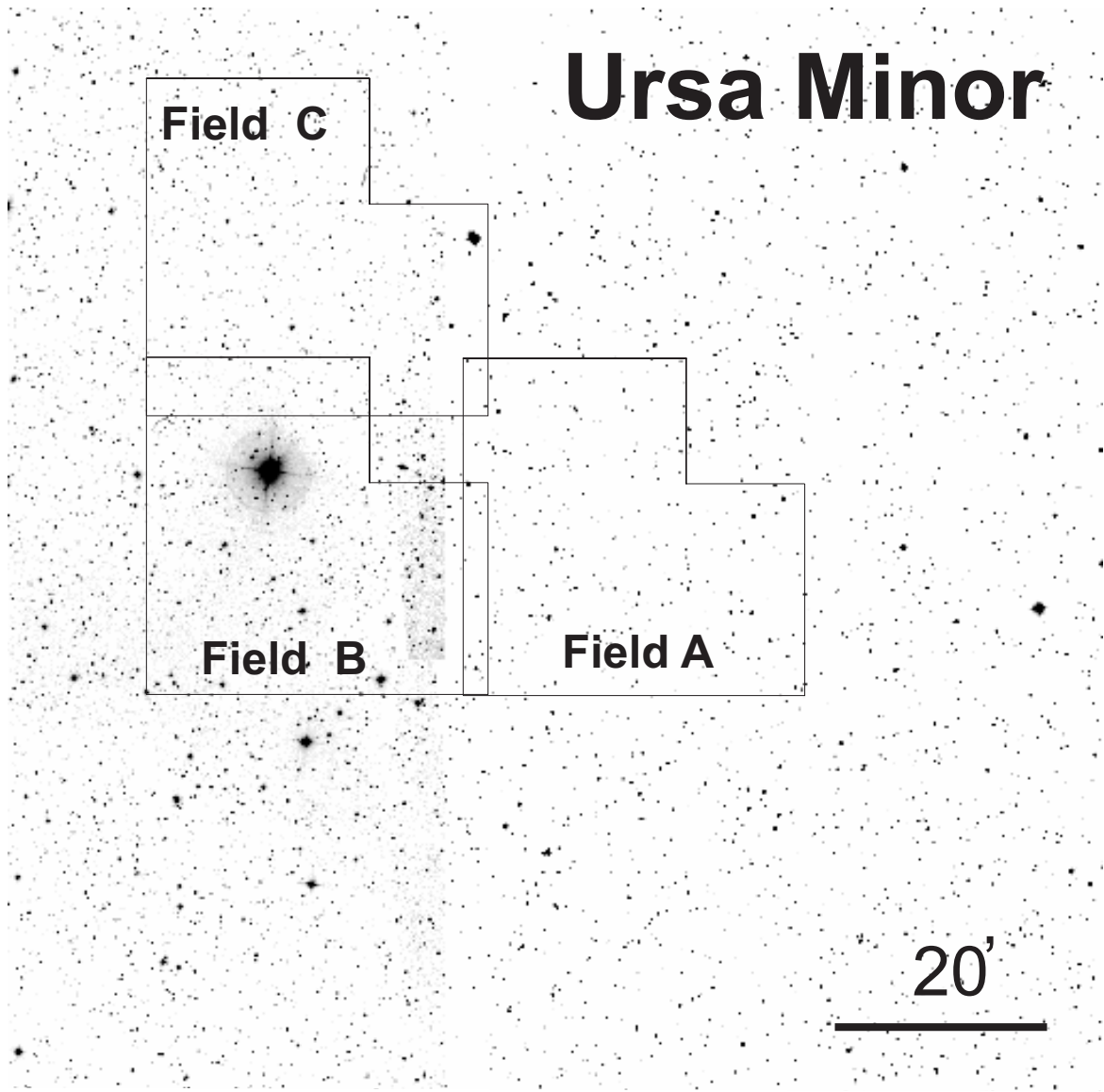,width=16cm}}
\figcaption[ima_1.eps]{Digitized Sky Survey image of the Ursa Minor
region. The observed fields are over-plotted. The galaxy is centered about the
center of Field A. North is up, East is left.
\label{ima_1}}
\end{figure}

The observations of Ursa Minor were carried out in $B$, $V$, $R$ and $I$
Johnson-Cousins filters with the 2.5 m Isaac Newton Telescope (INT) at Roque
de los Muchachos Observatory in La Palma and, with the 0.8 m, IAC-80
telescope at Teide Observatory in Tenerife. In the INT we used the WFC
installed at the prime focus, that holds four $2048 \times 4096$ pixels EEV
chips. The scale is $0.33''$/pix, which provides a total field of about $35'
\times 35'$. Three fields were observed in $B$ and $R$ filters encompassing a region of about $1\grados\times 1\grados$, with a total
covered area of about $0.75$ deg$^2$. Table \ref{journal} provides a
summary of the observations. Figure \ref{ima_1} shows the observed
fields. The field containing the center of the galaxy (field A) was also
observed in $V$ and $I$. These data will be used to estimate the distance and
metallicity (see \S 3). Observations with the IAC-80 telescope were done to
obtain the standard transformations for these bands. A Thomson $1024 \times
1024$ chip was used with a scale of 0.43$''$/pix, covering a total area of
nearly $7.5' \times 7.5'$.

\placetable{journal}
\placefigure{ima_1}

The images were processed in the usual way. Bias and flat-field corrections
were done with IRAF. DAOPHOT and ALLSTAR (Stetson 1994) were used to obtain
the photometry of the resolved stars. Transformation into the standard
photometric system requires several observations of standard star fields in
such a way that standards are measured in all four chips of the WFC. In
practice, we observed three fields from Landolt (1992) in the four chips and used them to
calculate relative photometric transformations from each chip to the central
one (chip 4). Then, a larger number of standard star fields were measured in
chip 4 during the observing run to calculate both nightly atmospheric
extinctions and the general transformation into the standard Johnson--Cousins
system. In total, 170 measurements of 38 standards were made during the
observing run of May 1998. Summarizing, the transformations from each chip to
chip 4 are:

\begin{equation}
(b_4-b_1)=0.329+0.02(B-R); ~~~~\sigma=0.005,
\end{equation}
\begin{equation}
(r_4-r_1)=0.837-0.307(B-R); ~~~~\sigma=0.010,
\end{equation}
\begin{equation}
(b_4-b_2)=0.443-0.006(B-R); ~~~~\sigma=0.007,
\end{equation}
\begin{equation}
(r_4-r_2)=0.833-0.268(B-R); ~~~~\sigma=0.008,
\end{equation}
\begin{equation}
(b_4-b_3)=0.367+0.021(B-R); ~~~~\sigma=0.008,
\end{equation}
\begin{equation}
(r_4-r_3)=0.739-0.233(B-R); ~~~~\sigma=0.008,
\end{equation}

\noindent  where sub-indices refer to chips, lower-case letters stand for
instrumental magnitudes, and capital letters for Johnson-Cousins magnitudes.
The $\sigma$ values correspond to the fit dispersion at the barycenter of
point distribution; hence they put lower limits to the zero-point
errors. 

Transformations from chip 4 instrumental magnitudes measured at the top of
the atmosphere into Johnson-Cousins magnitudes are given by:

\begin{equation}
(B-b_4)=24.553-0.061(B-R); ~~~~\sigma=0.009,
\end{equation}
\begin{equation}
(R-r_4)=24.393+0.257(B-R); ~~~~\sigma=0.010,
\end{equation}

\noindent where, as before, lower-case letters refer to extinction-free,
instrumental magnitudes and capital letters to Johnson--Cousins magnitudes.

The central field (field A in Fig. \ref{ima_1}) of Ursa Minor was also
observed with the INT in $V$ and $I$ in June 1999. This run was not
photometric, for which the IAC-80 telescope was used in May 2001 to obtain
the photometrical calibration. Fifteen standards from Landolt (1992) and
several secondary standards defined in the central region of Ursa Minor
(covered by chip 4 in the INT observations) were observed to this aim with
the IAC-80. The resulting transformation equations for the IAC-80 were:

\begin{equation}
(V-v_{\rm IAC80})=21.110+0.055(V-I); ~~~~\sigma=0.009
\end{equation}
\begin{equation}
(I-i_{\rm IAC80})=21.064+0.076(V-I); ~~~~\sigma=0.006
\end{equation}

\noindent while the final transformation for the INT observations (chip 4) resulted:

\begin{equation}
(V-v_{\rm 4})=24.13+0.04(V-I); ~~~~\sigma=0.02,
\end{equation}
\begin{equation}
(I-i_{\rm 4})=24.21+0.02(V-I); ~~~~\sigma=0.02,
\end{equation}

\noindent As formerly, lower-case letters refer to instrumental
magnitudes and capital letters to Johnson--Cousins magnitudes. 

Finally, dispersions of the extinctions for each night are about
$\sigma=0.01$ for each filter. Aperture corrections were obtained using a set
of $\sim 200$ isolated, bright stars in the Ursa Minor frames, dispersions
being of the order of $\sigma=0.01$. Putting all the errors together, the
total zero-point error of our photometry can be estimated to be about
$\sigma=0.02$ for all bands.

\begin{figure}
\centerline{\psfig{figure=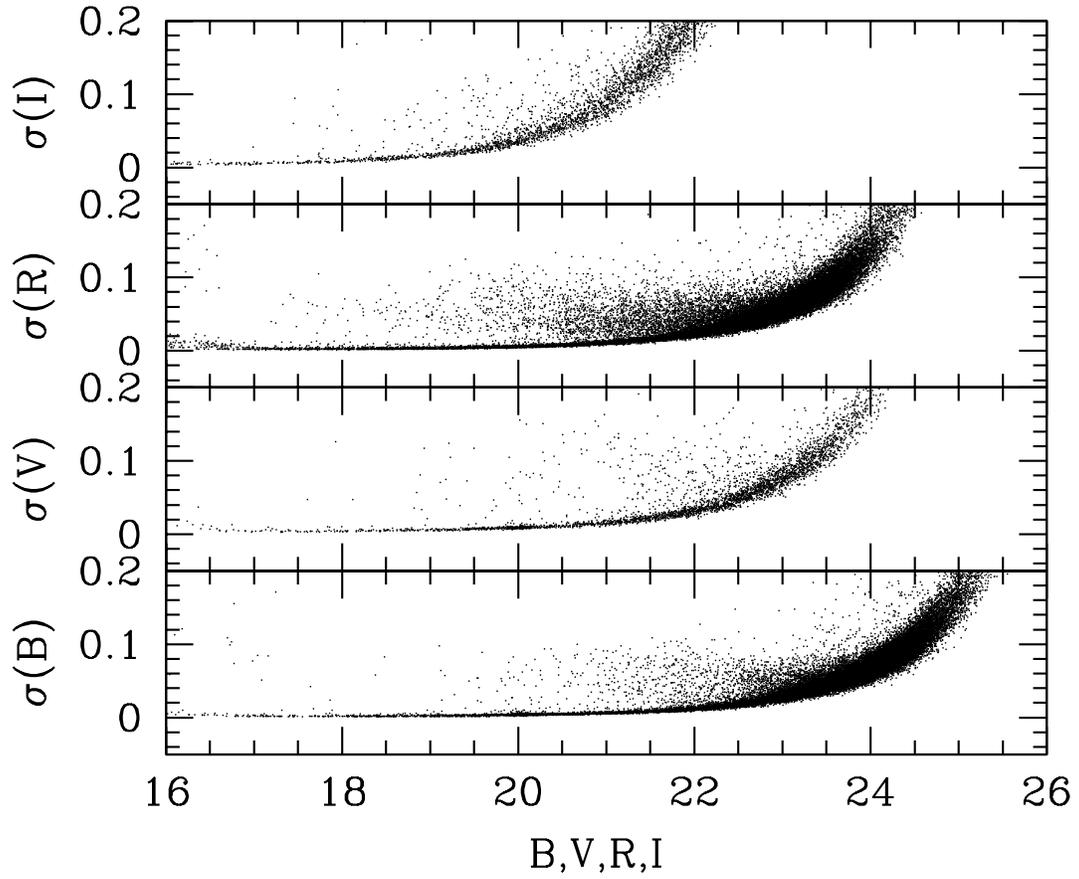,width=16cm}}
\figcaption[residuals.eps]{Residuals of the PSF-to-star fitting as a function
of magnitude provided by ALLSTAR for the stars resolved in the Ursa Minor
frames.
\label{residuals}}
\end{figure}

In addition to these errors, ALLSTAR provides, for each star, with the
dispersion, $\sigma$, of the PSF fitting. Generally, these residuals do not
reproduce the external errors of the photometry (see Aparicio, \& Gallart 1995 and Gallart,
Aparicio, \& V\'{\i}lchez 1996), but they are an indication of the internal
accuracy of the photometry. Figure \ref{residuals} shows these residuals as a
function of magnitude for each star. For the final photometric list we
selected stars with $\sigma<0.25$.

\placefigure{residuals}

\begin{figure}
\centerline{\psfig{figure=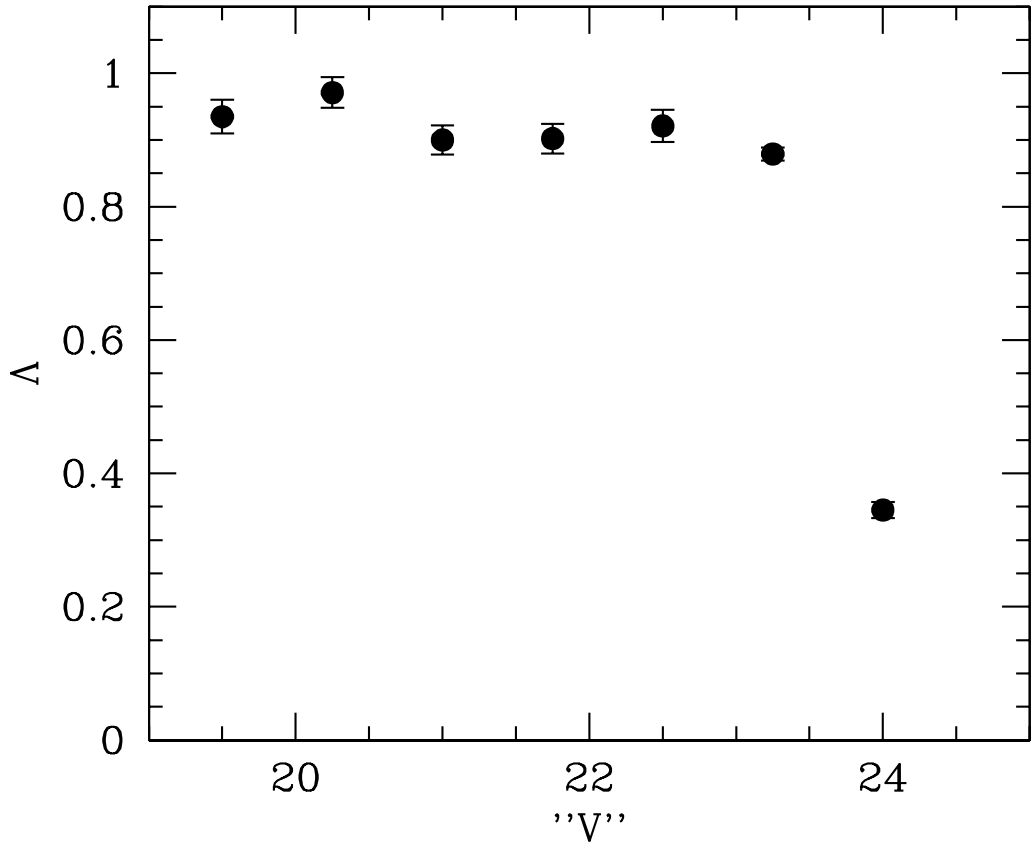,width=16cm}}
\figcaption[crow.eps]{The completeness factor as a function of $''V''$ for
the central chip of field A (see Fig. \ref{ima_1}). $"V"$ is defined as
$"V"=(B+R)/2$.
\label{crow}}
\end{figure}

Artificial-star tests have been performed in the usual way (Stetson \& Harris
1988; see also Aparicio \& Gallart 1995) to obtain completeness factors as a
function of magnitude. In short, 13~000 artificial stars were added to the
$B$ and $R$ images of Ursa Minor. The central chip (chip 4) of central field
was used. These stars were distributed in a grid covering the whole image. To
avoid overcrowding effects, they were separated about $8\farcs5$ from each
other. Colors and magnitudes chosen for artificial stars were obtained from
the synthetic CMD (computed as described in Aparicio \& Gallart 1995) of an
old stellar population with similar metallicity as Ursa Minor. The resulting
completeness factors are shown in Figure \ref{crow}. Completeness picks at
$''V''\simeq 23.5$. But about 5\% to 10\% of stars are lost at any brighter
magnitude. These stars are those happening to lie close to very bright,
saturated stars. 

\placefigure{crow}

\section{The color--magnitude diagram \label{cmds}}

\begin{figure}
\centerline{\psfig{figure=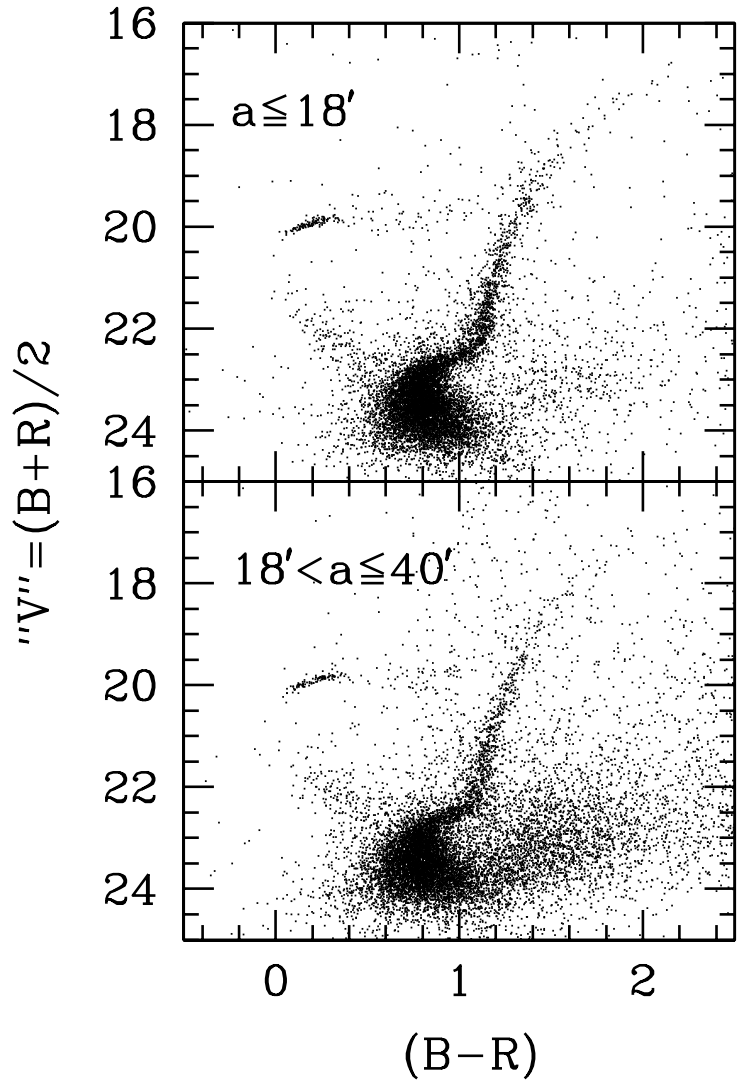,width=20cm}} 
\figcaption[cmd_br.eps]{The
CMD of the inner and outer regions Ursa Minor. Upper panel: stars within
$a\leq 18'$. Lower panel: Stars within $18'<a\leq40'$. $"V"$ is defined as
$"V"=(B+R)/2$.
\label{cmd_br}}
\end{figure}

\begin{figure}
\centerline{\psfig{figure=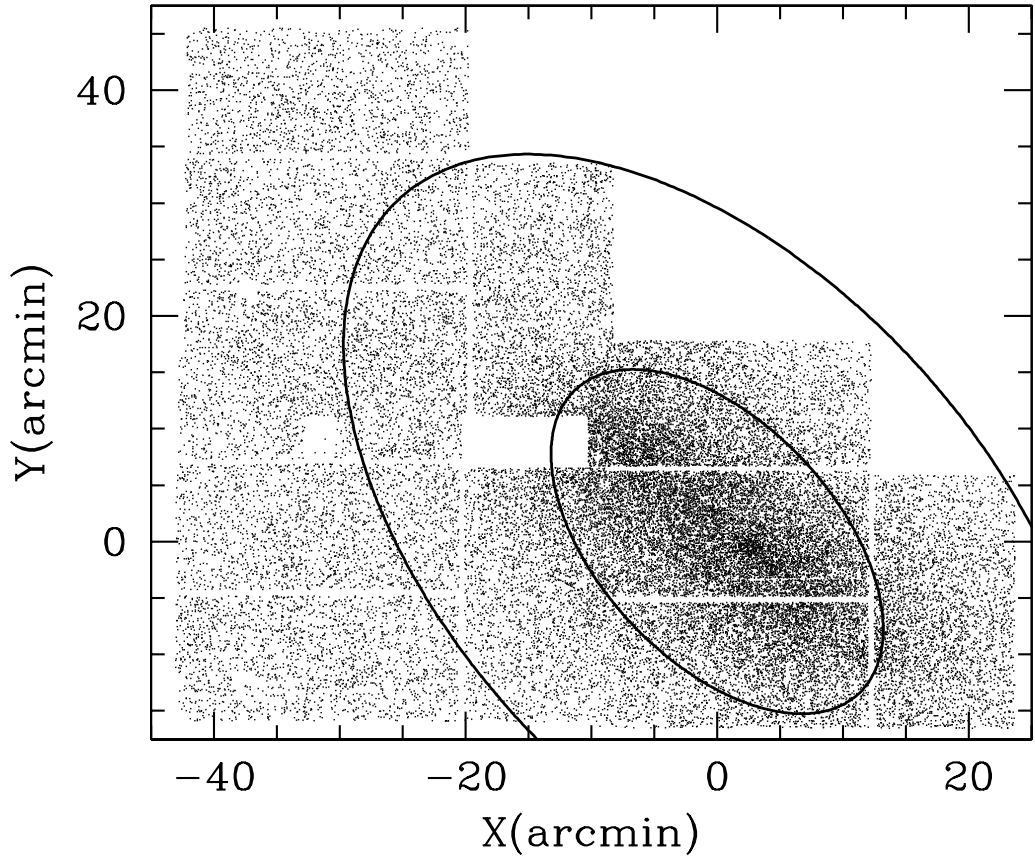,width=16cm}}
\figcaption[cmd_eli.eps]{Spatial distribution of the resolved stars together
with the ellipses used to spatially divide the galaxy. Ellipses semi-major
axis are $a=18'$ and $a=40'$. 
\label{david}}
\end{figure}

Figure \ref{cmd_br} shows the [$(B-R)$,$"V"$] CMD of the inner ($a\leq18'$)
and outer ($18'<a\leq 40'$) regions of Ursa Minor ($"V"$ is defined as
$"V"=(B+R)/2$). It must be noted that stars associated to Ursa Minor exist at
galactocentric distances larger than $40'$. We restrict our plot to stars
within this distance to limit foreground and background contamination
(see Mart\'\i nez-Delgado \etal 2001 for more details, including CMDs for
outermost regions). Figure \ref{david} shows a map of the resolved stars with
the ellipses limiting the inner and outer regions over-plotted. The galaxy
center, the ellipses ellipticity ($1-b/a=0.44$) and the position angle
($53\grados$) have been chosen after Irwin \& Hatzidimitriou (1995).

\placefigure{cmd_br}
\placefigure{david}

The two CMDs shown in Fig. \ref{cmd_br} are qualitatively similar. The most
noticeably features are the faint, main turn-off at $''V''\simeq 23$,
$(B-R)\simeq 0.8$; the blue-plume extending above it and reaching
$``V''\simeq 21.2$; a well populated, globular cluster-like, narrow red giant
branch (RGB), and a predominantly blue HB, including several stars in the
RR-Lyr instability strip. All this suggest an old, low metallicity stellar
population and a small metallicity dispersion. But the blue-plume requires
more attention because it could be made of blue-stragglers (BS) but it could
also be the trace of an intermediate-age population. It will be discussed in
\S\ref{bs}. 

\begin{figure}
\centerline{\psfig{figure=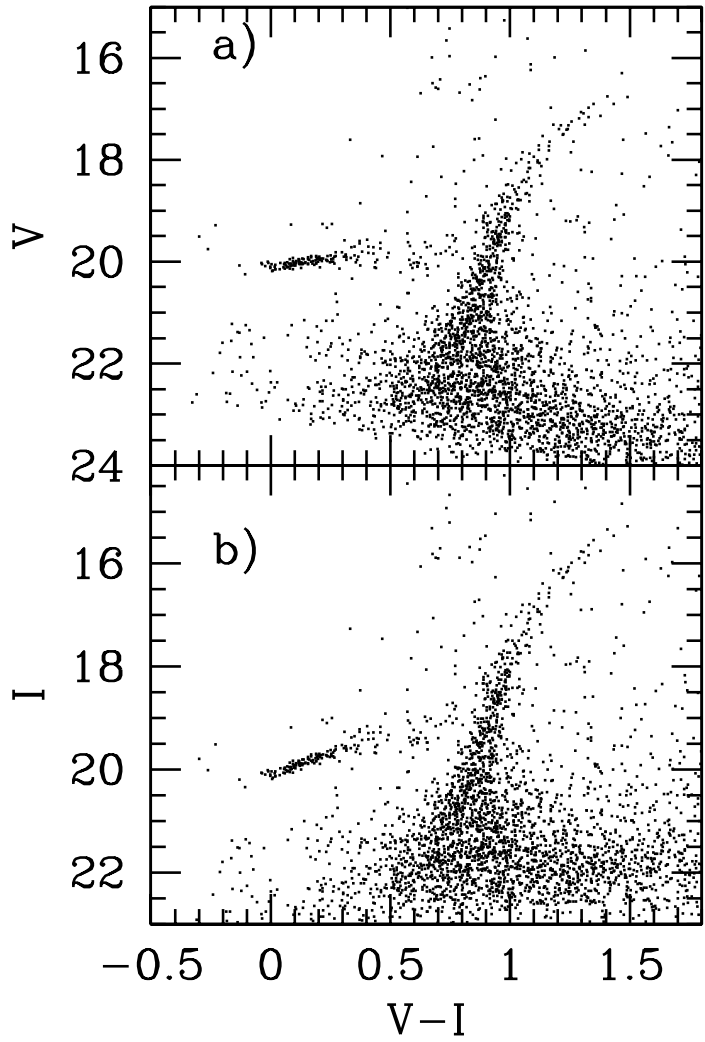,width=20cm}}
\figcaption[cmd_v.eps]{$[(V-I),V]$ and $[(V-I),I]$ CMDs corresponding to the
central chip of the WFC in Field A (see Fig. \protect{\ref{ima_1}}).
\label{cmd_vi}}
\end{figure}

Figure \ref{cmd_vi} shows the [$V-I$,$V$] and [$V-I$,$I$] CMDs of stars in
the central chip of field A. Although shallower and less populated than the
[$(B-R)$,$"V"$] CMD, they are better suited for the estimation of distance
and metallicity and will be used for this purpose in the following
paragraphs.

\placefigure{cmd_vi}

\subsection{The distance}

The distance to Ursa Minor can be calculated from the HB magnitude at the
color of the RR Lyrae variables. A relation between $[Fe/H]$ and $M_V$ of RR
Lyrae must be assumed to this purpose. However, Demarque \etal (2000) have
shown that such relation is not universal and depends on the HB
morphology. To overcome this problem, we have compared Ursa Minor HB with
that of several globular clusters with similar metallicity and HB morphology
as the galaxy. The distance to Ursa Minor can then be obtained from
the difference in magnitude between cluster and galaxy HBs and the cluster
distances. 

For this purpose, we firstly need an estimate of Ursa Minor metallicity and
of its HB $V$ magnitude. The metallicity has been estimated using the color
of the RGB and the Carretta \& Gratton (1997; CG from here on) and Zinn \&
West (1984; ZW from here on) scales (see \S \ref{metal}). The first one,
$[Fe/H]=-1.75$, will be adopted here, since the metallicities of the globular
clusters used for comparison are given in the CG scale. 

The Ursa Minor HB $V$ magnitude must be a reddening free value, which we will
denote $V^{\rm UMi}_{\rm HB,0}$, obtained approximately for the center of the
RR-Lyr instability strip, i.e., at about $(V-I)=0.4$. The HB, $''V''$
magnitudes must be firstly transformed into usual $V$ Johnson
magnitudes. This has been done from the same Padua stellar evolution models
(Bertelli et al. 1994) used in \S\ref{sfh}. The average difference for the
interval $0.2\leq (B-R)\leq 0.6$ is $V-''V''=-0.031\pm 0.002$. Then, an
empirically defined fiducial HB taken from Rosenberg \etal (1999) has been
fitted to the Ursa Minor HB. It produces $V^{\rm UMi}_{\rm
HB}=19.84\pm0.07$. Furthermore, using the IR dust maps by Schlegel,
Finkbeiner \& Davis (1998), a reddening of $E(B-V)=0.034$ is obtained for
Ursa Minor. Assuming the relations by Cardelli, Clayton, \& Mathis (1989),
this corresponds to an extinction $A_V=0.104$, from which $V^{\rm UMi}_{\rm
HB,0}=19.74\pm0.07$ results.
 
\placetable{cgg}

The clusters selected for the HB comparison are listed in Table
\ref{cgg}. They have been taken from the IAC-Padua Catalogue of Galactic
Globular Clusters (CGGC) (Rosenberg \etal (2000a,b). Column 1 gives the
cluster name. Column 2 lists the metallicity in the CG scale. The
reddening-free distance moduli and HB magnitudes are given in columns 3 and
4. These values are estimated from main-sequence fitting to Hipparcos
subdwarfs (Reid 1999 and references therein). Although clusters with similar
metallicity as Ursa Minor have been chosen, small differences are still
present. To avoid the effects of these differences, the globular cluster HB
magnitudes have been transformed to those corresponding to the galaxy
metallicity using $V^{\rm gc}_{\rm HB,0,[Fe/H]}=V^{\rm gc}_{\rm
HB,0}+0.18([Fe/H]_{UMi}-[Fe/H]_{gc})$. The resulting magnitudes are given in
column 5. Column 6 lists the reddening-free magnitude differences between
Ursa Minor and cluster HBs: $\Delta V_{\rm HB,0}=V^{\rm UMi}_{\rm
HB,0}-V^{\rm gc}_{\rm HB,0[Fe/H]}$. Finally, column 7 gives the distance
moduli obtained for Ursa Minor as $(m-M)^{\rm UMi}_0=(m-M)^{\rm
gc}_0+\Delta_{\rm HB,0}$.

The average of values listed in table \ref{cgg}, column 7 is
$(m-M)_0=19.4\pm0.1$, corresponding to $d=76\pm4$ Kpc. This value is larger
than the value $(m-M)_0=19.24\pm0.24$ given by Nemec \etal (1988) using RR
Lyrae periods. Comparing Ursa Minor and M92 fiducial sequences, Miguell \&
Burke (1999) and Olzewski \& Aaronson (1985) derived distance moduli of
$(m-M)_0=19.18\pm0.12$ and $(m-M)_0=19.0\pm0.1$, respectively, also smaller
than our result.


\subsection{The metallicity \label{metal}}

The metallicity of Ursa Minor has been measured spectroscopically by Shetrone
\etal (2001), who provided a mean value of $[Fe/H]=-1.90\pm0.11$. Previously
Olzewski \& Aaronson (1985) photometrically estimated it to be about
$[Fe/H]=-2.2$, assuming that Ursa Minor has the same metallicity and age as
M92. Here, we will provide a new photometric estimate, based on the color of
the RGB. This is a frequently used estimator because photometrically
observing and measuring the RGB is a relatively simple task. However, it is
clear that this method can not compete with spectroscopic determinations and
recently, Gallart et al. (2001) have shown that these RGB-estimates can be
wrong by far. In any case, the metallicity we will derive will, at least, be
useful for comparison with results for other galaxies existing in the
literature obtained with the same method.

To estimate the metallicity, we have used the relations provided by Saviane
\etal (2000), based on the position of the RGB of several globular clusters
of the IAC-Padua CGGC (Rosenberg \etal 2000a,b). We will use the calibration
based in the $(V-I)_{I=-3}$ parameter; i.e., the $(V-I)$ color index of the
RGB at $M_I=-3$. From figure \ref{cmd_vi} we estimate $(V-I)_{I=-3}=1.17\pm
0.03$. The resulting metallicity is $[Fe/H]=-1.75\pm 0.20$ if the CG
metallicity scale is used, or $[Fe/H]=-2.01\pm 0.14$ if the ZW metallicity
scale is used. Considering both values, we adopt $[Fe/H]=-1.9\pm 0.2$, which
agrees with the spectroscopic value.

\begin{figure}
\centerline{\psfig{figure=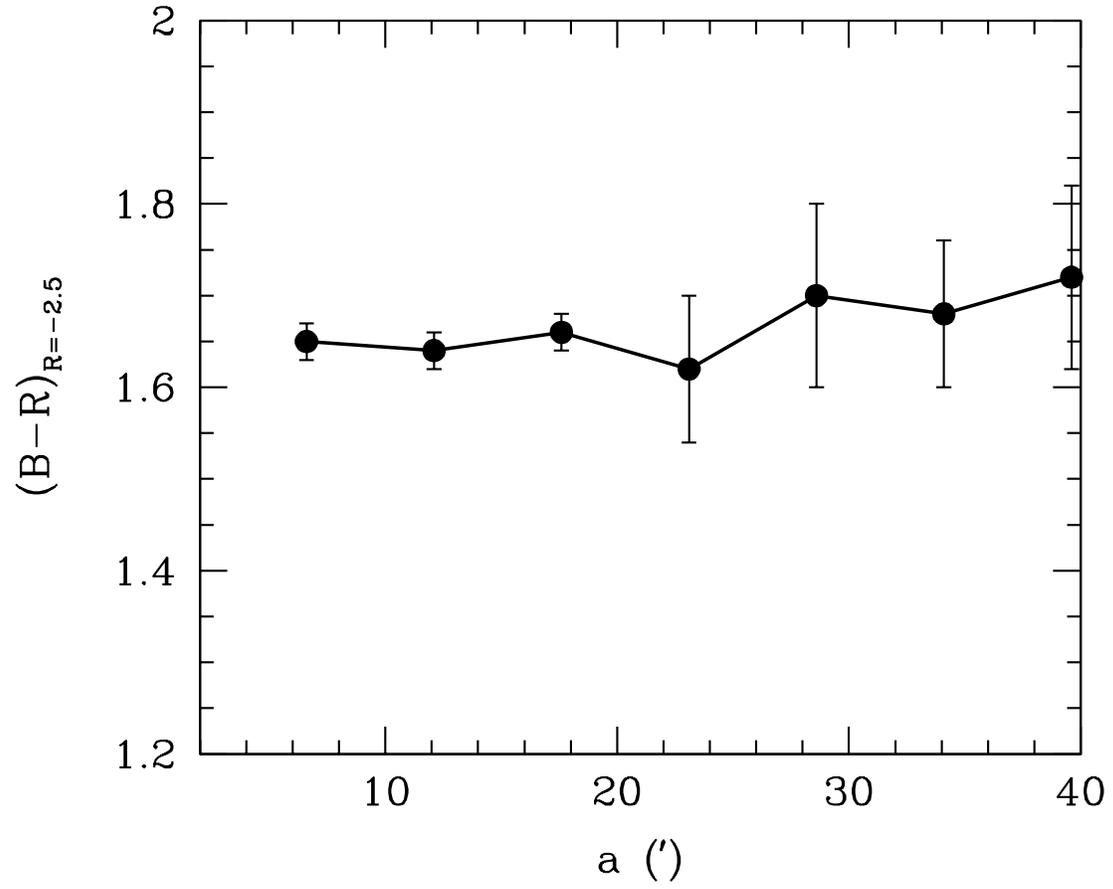,width=16cm}}
\figcaption[met_grad.eps]{$(B-R)_{R=-2.5}$, the color of the RGB measured at
$M_R=-2.5$, of Ursa Minor stars for elliptical annuli of increasing semimajor
axis $a$.
\label{met_grad}}
\end{figure}

The color of the RGB can be also used to test whether galactocentric
metallicity gradients exist in Ursa Minor, under the reasonable assumption
that age gradients, if any, have negligible effects on the RGB color. To this
purpose, the galaxy has been divided in several concentric elliptical annuli
and the [$(B-R)$,$R$] CMDs have been plotted for them. The $BR$ plane is used
now because of the larger spatial coverage made in the observations with
these filters. Hyperbolic polynomial fits to points in the interval $-1.5\leq
R\leq -3.2$ were used to obtain a fiducial RGB for each CMD. The $(B-R)$ at
$M_R=-2.5$ ($(B-R)_{R=-2.5}$) has then been used as an estimator of the
metallicity (Aparicio \etal 2001). It was measured for
each fit and the results are plotted in Figure \ref{met_grad}. The error bars
show the internal dispersions of points about the polynomial fits. Inspection
of this figure reveals that no metallicity gradient can be deduced to exist
in Ursa Minor.
   
\placefigure{met_grad}

\section{The Star Formation History\label{sfh}}

The SFH of a galaxy can be derived in detail from a deep CMD through
comparison with synthetic CMDs (Aparicio 2001). To obtain the SFH of Ursa
Minor, we have used here the {\it partial model} method, introduced by
Aparicio, Gallart, \& Bertelli (1997). We have closely followed the
prescriptions and criteria used by Aparicio \etal (2001), where this method
was applied to derive the SFH of the Draco dSph. Details on the computation
of synthetic CMDs and on the SFH computation method itself can be found in
Gallart (1999) as well as in Aparicio \etal (1997), as well as in
Aparicio (2001) where different methods are reviewed.

The SFH can be considered a function of several variables (time, stellar
mass, etc). A typical approach is assuming it composed by the star formation
rate (SFR) function, $\psi(t)$ and the chemical enrichment law (CEL),
$Z(t)$. But the initial mass function (IMF), $\phi(m)$ and a function
accounting for the fraction and mass distribution of binary stars,
$\beta(f,q)$ are also relevant. The most powerful methods now in use are
able to solve for $\psi(t)$ and $Z(t)$ simultaneously and treat $\phi(m)$ and
$\beta(f,q)$ as inputs (see Aparicio 2001 and references therein). In our
case, to simplify and having Ursa Minor low metallicity and low metallicity
dispersion, we have fixed $Z(t)$ and have solved for $\psi(t)$ alone. We have
adopted the Kroupa, Tout, \& Gilmore (1993) IMF and have neglected the
contribution of binary stars.

The metallicity given in \S \ref{metal}, $[Fe/H]=-1.9\pm 0.11$ (Shetrone \etal
2001), corresponds to $Z=(2.5\pm0.5)\times 10^{-4}$. We have adopted a
constant metallicity law with this value, but allowing some dispersion, to
compute the synthetic stars. According to it, synthetic stars have
metallicities randomly distributed in the interval $2\times 10^{-4}\leq Z\leq
3\times 10^{-4}$, independently of age. This produces a RGB compatible in
color and wideness with Ursa Minor's.

The only remaining function to complete the SFH of Ursa Minor is the SFR,
$\psi(t)$. It has been solved using the aforementioned {\it partial model}
method. To apply it, a single synthetic CMD is computed, embracing the full
possible age interval (e.g. from 15 Gyr to 0 Gyr). This is divided in
several partial models, each containing a stellar population with ages in a
narrow interval. The Padua stellar evolution library (Bertelli \etal 1994) is
used to this purpose. The basic idea is that any SFR can be simulated as a
linear combination of partial models.

The solution for the SFR must have associated to it a CMD compatible with the
observed one. To find such solution, a number of boxes are
defined in the CMDs (observed and partial models) and the number of stars are
counted inside them. The boxes are defined in such a way they sample stellar
evolutionary phases providing information about different ages. Let us call
$N_j^o$ the number of stars in box $j$ in the observed CMD and $N_{ji}^m$ the
number of stars in box $j$ of partial model $i$ (the partial model covering
the $i$-th age interval). Using this, the distribution of stars in boxes of an
arbitrary SFR can be obtained from a linear combination of the $N_{ji}^m$, by
\begin{equation}
N_j^m=k\sum_i\alpha_iN_{ji}^m.
\end{equation}
\noindent The corresponding SFR can be written as
\begin{equation}
\psi(t)=k\sum_i\alpha_i\psi_p\Delta_i(t),
\end{equation}
\noindent where $\alpha_i$ are the linear combination coefficients, $k$ is a
scaling constant, and $\Delta_i(t)=1$ if $t$ is inside the interval
corresponding to partial model $i$, and $\Delta_i(t)=0$ otherwise. Finally,
the $\psi(t)$ having the best compatibility with the data can be obtained by
a $\chi^2$ fitting of $N_j^m$ to $N_j^o$, the $\alpha_i$ coefficients
being the free parameters. 

In practice, several choices of partial model distributions have been
tried. However, choices involving time resolutions for old stars of the order
of 1 Gyr are fluctuating in a similar way as shown by Olsen (1999),
indicating that data do not contain enough information for such small time
sampling. On the other hand, the SFR at intermediate ages is so low (if any)
that only an average estimate makes sense. In summary, after several trials,
the adopted solution is based on a three partial model decomposition
corresponding to the age intervals 2-10 Gyr, 10-13 Gyr and 13-15 Gyr.  The
lower limit (2 Gyr) is the age of an isochrone having the turn-off point at
$''V''\sim 1.5$, corresponding to the upper point of the Ursa Minor blue
plume. 

\begin{figure}
\centerline{\psfig{figure=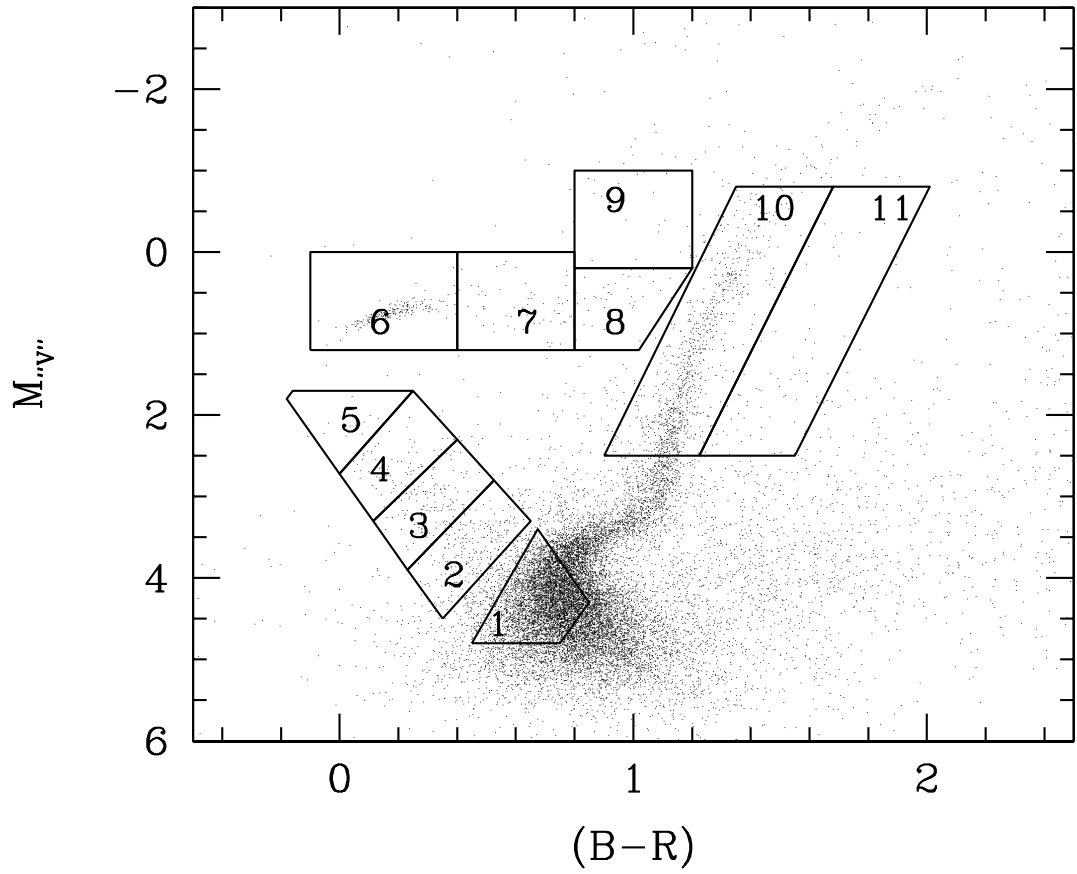,width=16cm}}
\figcaption[cmd_box.eps]{Boxes used for the study of Ursa Minor's SFH,
over-plotted on the CMD corresponding to the inner $18'$.
\label{cmd_box}}
\end{figure}

Moreover,11 boxes have been defined in the CMD, as shown in
Fig. \ref{cmd_box}, to characterize the distribution of stars. Box 1 samples
the old MS population. Boxes 2 to 5 sample the BP population. In this
analysis we will consider two possibilities in turn, namely that all the BP
stars are genuine MS stars (hence of intermediate-age) and that all them are
BS (hence old stars). We will discuss more about this in \S\ref{bs}. Box 6 to
8 cover the HB, while box 9 samples the region corresponding to intermediate
mass, core He-burning stars. Here, we will assume that age is the only
relevant second parameter to determine the HB morphology. However, it must be
kept in mind that the nature of the second parameter is not fully confirmed
and that the blue-ward extension of the HB depends on stellar evolution
parameters that are not well controlled. This could affect the morphology of
the SFR for the interval 10-15 Gyr, although the effects on the SFR integral
for that period would be very limited.

Box 10 samples the RGB, populated by old and intermediate-age stars born in
fact in the full, 2--15 Gyr interval considered here. For this reason, it
gives no information on age resolution, but it provides a strong constraint
to normalize the full SFR: independently of the age distribution, the
integrated SFR for the old and intermediate-age stars must be compatible with
it. Beside a few AGBs in the same age interval as the RGBs, box 10 includes a
number of foreground stars. This number is estimated from box 11 and simply
subtracted from the counts in box 10.

\placefigure{cmd_box}

The solution for the SFR is found in a global way, considering the number of
stars in all boxes. In practice, it is not reasonable to look for the best
solution (the linear coefficients best reproducing the distribution of stars
in the boxes of the observational CMD), but to search all the solutions
providing a stellar distribution compatible with the observed one within some
interval. The criterion used here is to adopt as good solutions all those
producing $chi^2_\nu$ (the reduced $\chi^2$) within $\chi^2_{\nu,{\rm
best}}+1$. 

\begin{figure}
\centerline{\psfig{figure=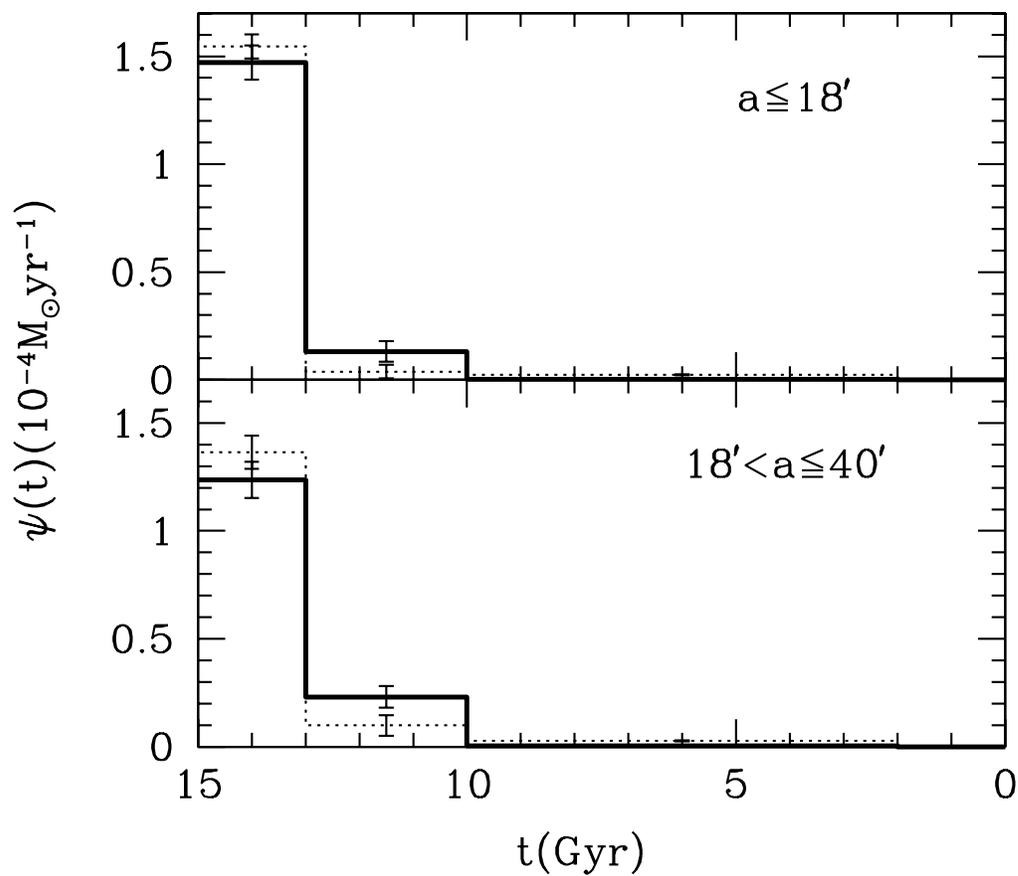,width=16cm}}
\figcaption[sfr_pm.eps]{The SFR as a function of time for the inner
($a\le18'$) and outer ($18'<a\le40'$) regions of Ursa Minor. Thick, solid
lines show the solutions obtained assuming that the BP stars are BS. Thin,
dotted lines, show the solutions obtained assuming that the BP stars are
genuine MS stars. Error bars represent the solution dispersions for each age
interval.
\label{sfr_pm}}
\end{figure}

\placefigure{sfr_pm}

The large spatial coverage of our data offers the opportunity of comparing
the stellar population in different regions of the galaxy. To do so, we have
computed $\psi(t)$ in the inner ($a\leq 18'$) and in the outer ($18'<a\leq
40'$) regions defined in \S\ref{cmds}. These values approximately correspond
to the core and tidal radii (e.g. Irwin \& Hatzidimitriou 1995 find
$r_c=15\farcm8$, $r_t=50\farcm6$ and Kleyna \etal 1998 obtain $r_t=34'$). The
results are shown in Fig. \ref{sfr_pm}. Error bars represent the dispersion
of the accepted solutions.

As in the case of the Draco dSph (Aparicio \etal 2001), the solutions are
very similar in both regions. However, a marginally significant difference
could exist in the fraction of stars in the $10-13$ Gyr interval relative to
the $13-15$ Gyr one, in the sense that the former could be relatively more
frequent in the outer region of the galaxy. Since the difference is at
the error bar level we do not further discuss on it. 

On the other hand, the solution is largely dominated by very old stars. Some
90\% of the stars were formed in an initial, main burst lasting from 15 to 13
Gyr ago and at least 95\% of them were formed before 10 Gyr ago. After that,
two possibilities arise. If BP stars are normal MS stars, then star formation
has gone on at a low rate until recent epochs. Alternatively, if BP stars are
BS, the star formation has been zero or negligible after 10 Gyr ago. For the
reasons discussed below (\S\ref{bs}) we adopt the second possibility as the
most reliable, which makes Ursa Minor the only Milky Way satellite lacking an
intermediate age stellar population.

The obtained SFR has been used to calculate the total mass in stars and
stellar remnants ($M_\star$) within $18'$ and within $40'$. Both values are
quoted in Table \ref{densi_par} and the second is used to calculate the dark
matter fraction, $\kappa$.

The SFR has been also computed by Hern\'andez \etal (2000) (see also
Valls-Gabaud et al. 2000) for the central region of Ursa Minor using HST
data. They obtained a solution qualitatively very similar to ours: a major
star formation event at an early epoch forming all or almost all the stars in
the galaxy. However, Hern\'andez \etal (2000) find a maximum value of
$6\times10^{-4}$M$_{\odot}$yr$^{-1}$ at about 14 Gyr and 0 for ages $<10$ Gyr
while Valls-Gabaud et al. (2000), $3\times 10^{-4}$M$_\odot$yr$^{-1}$, also
at $\sim 14$ Gyr. These authors calibrate their SFR scale by normalization to
account for the total luminosity of Ursa Minor. To compare with them we have
to sum the solutions for our inner and outer regions. We obtain a maximum
SFR of $2.7\times 10^{-4}$M$_\odot$yr$^{-1}$ for the period 15-13 Gyr, quite
similar to the Valls-Gabaud et al. (2000) value. It is important to stress
the compatibility of both solutions, which is more noticeable considering the
different approaches used by these authors and us.

\section{On the nature of blue stars\label{bs}}

To derive the SFH of Ursa Minor, we have considered in \S\ref{sfh} two
possibilities about the nature of BP stars, namely that they are normal,
intermediate-age, MS stars or BS. Although it is probably impossible to
conclusively ascertain what is the right possibility, a few tests can be done
to check what of them is the most reliable. We will discuss in turn the
following points, that should cast light on the problem: (i) the amount of
stars in the RC region of the CMD; (ii) the relative number of BP stars;
(iii) their spatial distribution, and (iv) the availability of gas in Ursa
Minor.

\subsection {Stars in the RC region}

If BP stars would be genuine, intermediate-age, MS stars, they should have a
counterpart in the red-clump region of the CMD, sampled by box 9 (Figure
\ref{cmd_box}). However, this criterion is not sensitive enough: only 2-3
stars of this kind are expected to be found in this box, while some 20 to 22
old, AGB stars evolving upwards from the HB are expected to lie in the same
box. This must be added to the fact that some foreground and background
contamination is expected to affect this box, making the star counts in it
more uncertain.

\subsection {The relative number of blue stars}

The number of BS stars relative to a reference population, namely the
HB stars, is used in the study of the frequency and distribution of BS
in globular clusters and field stars. We have computed this number for
Ursa Minor. Only the inner $18'$ have been used to limit the effects
of background and foreground contamination. The resulting number of BS
relative to HB stars is $F_{BS}^{HB}=1.8$. This can be compared with
the results obtained for globular clusters by Piotto et al. (2002) and
for Milky Way halo field, blue, metal-poor stars (Preston \& Sneden
2000). The former obtain values in the range $0.1\leq F_{BS}^{HB}\leq
1.0$ for most globular clusters in their sample, larger values
corresponding to less concentrated clusters. The only two exceptions
have $F_{BS}^{HB}$ between 2 and 3. Preston \& Sneden (2000) obtain
$F_{BS}^{HB}=4.4$. This scenario points to the BS possibly originating
in close binary stars, that would be more efficiently destroyed in
high density globular clusters. This would account for the higher
values of $F_{BS}^{HB}$ in lower density environments. Indeed, the
value found here for Ursa Minor, which is intermediate between those of
globular clusters and that of the Milky Way halo field, would be compatible
with this scenario and would agree with the BP stars in Ursa Minor
being BS stars.

\subsection {The spatial distribution of blue stars}

\begin{figure}
\centerline{\psfig{figure=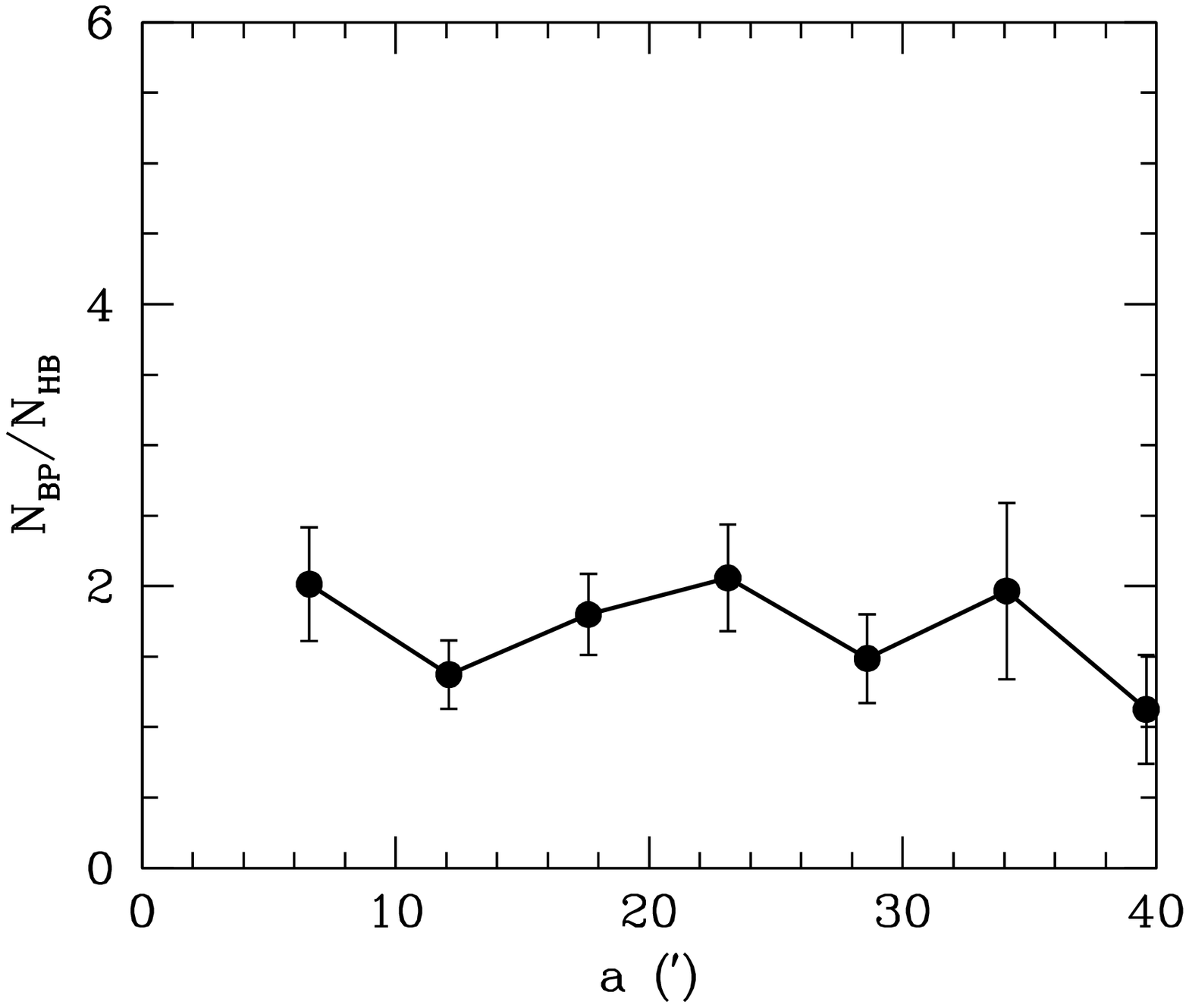,width=16cm}}
\figcaption[bsrad.eps]{The radial distribution of the BP stars relative to
HB. Boxes 2 to 5 and 6 to 8 (as defined in Figure \protect\ref{cmd_box}) have
been respectively used to sample the BP and HB stars.
\label{bsrad}}
\end{figure}

A further test of the nature of the BP stars is based on their radial
distribution. The gas coming from old, dying stars would have a radial
distribution similar to that of these stars. But the fact that a critical
density of gas is required to form new stars implies that these should
preferentially form in the inner part of the galaxy (see, for example, the
cases of Phoenix and NGC 185 in Mart\'\i nez-Delgado et al. 1999a, b,
respectively). Figure \ref{bsrad} shows that the distribution of
$F_{BS}^{HB}$ as a function of galactocentric distance in Ursa Minor is flat,
indicating that the BP population is strongly related to the old stellar
population and that it is very likely formed by BS stars.

\placefigure{bsrad}

\subsection {Gas and intermediate-age stars in Ursa Minor\label{bs_gas}}

Producing an intermediate-age population implies a mechanism to generate and
conserve gas. Deep observations have failed to detect gas in Ursa Minor
(Young 1999, 2000), even for large distances from the center of the galaxy
(Blitz \& Robishaw 2000). However, the gas could come, at a low rate, from
early generation, dying stars. This could eventually allow small, short and
recursive star formation bursts that, smoothed and averaged, would account
for a low rate star formation at intermediate ages. The rate at which this
material is returned to the ISM can be estimated from the integration of the
SFR through

\begin{equation}
R(t)=\int_{m_{l}}^{m_{u}}[m-p(m)]\phi(m)\psi[t-\tau(m)]dm
\end{equation}

\noindent where $m_l$ and $m_u$ are lower and upper integration limits for
the stellar mass, $p(m)$ is the mass of the stellar remnant of a star of
initial mass $m$, and $\tau(m)$ is the life time of a star of mass $m$. Note
that we are using time increasing toward the past, with the present-day value
being $0$. We assume that all the gas returned to the interstellar medium
from SNe explosions escapes the galaxy and we will estimate only the gas
coming from stellar winds of intermediate and low mass stars in the inner
Ursa Minor region, where density is larger. From a linear fit to Vassiliadis
\& Wood's (1993) results for stars of metallicity $Z=0.004$ (the lowest they
compute) in the mass interval from 0.89 M$_\odot$ to 5.0 M$_\odot$, we obtain
$p(m)=0.489+0.093\times m_i$, where $m_i$ is the initial stellar mass. Using
the SFR obtained in \S\ref{sfh} for the inner region ($a\leq 18'$) and for
the case in which BP stars are assumed to be intermediate-age MS stars; the
Kroupa \etal's (1993) IMF; $m_l=0.89$M$_\odot$, and $m_u=5.0$, we obtain in
average $\bar R(t)=1.3\times10^{-5}$ M$_{\odot}$yr$^{-1}$ for $2$ Gyr$<t<13$
Gyr (inner $18'$). This value is five times the averaged SFR
obtained in \S\ref{sfh} for that period, $\bar\psi_{13-2}=2.5\times 10^{-6}$
M$_{\odot}$yr$^{-1}$.

However, in a dwarf galaxy such as Ursa Minor, gas is expected to be
repeatedly removed through SNe explosions. Van den Bergh \& Tammann (1991)
estimated that, for a galaxy of the luminosity of Ursa Minor, the SNIa rate
is one per $10^7$ years. In this interval Ursa Minor (inner $18'$) would
accumulate some $130$ M$_{\odot}$ in all of gas from intermediate (if any)
and low mass dying stars. Under these conditions, it seems really difficult
to accumulate gas enough in any limited region of the galaxy to allow stars
to form at intermediate ages.

To see this more clearly it should be enough considering that, if gas from
intermediate and low mass dying stars would distribute across the galaxy as
the stars do, the central gas density would reach a value of some $4.4\times
10^{-3}$ M$_\odot$pc$^{-3}$ after 10 Gyr if not removed by SNe
explosions. This value corresponds to a particle density of $5\times 10^{-3}$
cm$^{-3}$, much smaller than values about $1000$ cm$^{-3}$, which are typical
lower thresholds for star forming regions (Shu, Adams, \& Lizano 1987).

\section{The SFH scenario in Ursa Minor\label{gas}}

In summary, the tests discussed in \S\ref{bs} support the idea that BP stars
in the Ursa Minor CMD are BS and that not a significant intermediate-age
stellar population is present in the galaxy. Consequently, the overall SFH
for Ursa Minor can be sketched in the following way. In a first phase,
lasting from about $15$ to $13$ Gyr ago with a further extension up to 10 Gyr
ago, most or all the stars now populating the galaxy were formed. Unused
gas would likely have been ejected or swept in the subsequent SNe
explosions. In a second phase, starting upon the end of the first one, dying
stars injected gas into the interstellar medium. Some of it could have been
used to form new stars but, most likely, SNIa explosions would have avoided
the accumulation of enough gas at any moment of this phase, preventing any
star formation at intermediate ages. In such a case, blue-plume stars in Ursa
Minor would be BS. The best solution for the SFH is shown by the full line in
Figure \ref{sfr_pm}. Alternatively, the maximum SFR compatible with the BP
stars being genuine, intermediate-age, MS stars is shown by a dotted line
in the same figure.

It must be mentioned that, in our analysis of Draco (Aparicio et al. 2001),
we assumed that the BP was formed by genuine MS stars. The 2-3 Gyr old burst
found in the SFH of Draco depends on this assumption and, for this reason, as
quoted in Aparicio et al. (2001), the value found for of the SFR in that
burst is to be considered an upper limit. But the burst also relies on the RC
and subgiant populations present in Draco's CMD, for which the existence of
an intermediate-age stellar population in Draco should be considered
real. Summarizing, although apparently similar at first glance, Ursa Minor
and Draco differ in two important properties: their dynamical structures,
with Ursa Minor showing a tidal tail (Martínez-Delgado et al. 2001) but Draco
lacking it (Aparicio et al. 2001) and Draco showing an intermediate-age
stellar population which is not present in Ursa Minor.

\section{Conclusions}

As a part of our project devoted to the study of the Ursa Minor dSph, the SFH
of the galaxy is presented in this paper. The analysis uses wide field
photometry, encompassing about $1\grados\times1\grados$ (total
covered area about $0.75$ deg$^2$) and the synthetic CMD technique for the
derivation of the SFH.

The resulting SFH shows that Ursa Minor hosts a predominantly old stellar
population. Very likely, virtually all the stars were formed before 10 Gyr
ago, and 90\% of them were formed before 13 Gyr ago. This picture shows Ursa
Minor as the Milky Way dSph satellite most resembling the original hypothesis
of Baade, namely that dSphs were pure old, globular cluster-like
systems. Nevertheless, Ursa Minor CMD shows a well populated blue-plume above
the main, old, MS turn-off. If these stars were genuine, MS stars, Ursa Minor
would have maintained a low star formation rate extending up to $\sim2$ Gyr
ago. However, (i) the relative number of BP to HB stars; (ii) the spatial
distribution of BP stars, and (iii) the gas availability play against such
possibility and strongly favors the BP stars being a BS population. In this
context, Ursa Minor would remain the only dSph, Milky Way satellite hosting a
pure old stellar population.

The SFH study has been done for two regions of the galaxy, namely the one
within the core radius and the one between this and the tidal radius. No
significant differences have been found between the stellar populations in
both regions except for the fact that the fraction of stars found in the
$10-13$ Gyr interval is marginally larger. 

Besides the SFH analysis, the distance of Ursa Minor has been also derived
from comparison of the galaxy HB magnitude with those of some globular
clusters. These were chosen to have the same metallicity and HB morphology
than the galaxy. We obtained a distance of $d=76\pm4$ Kpc. The metallicity
was also estimated from the RGB color to be $[Fe/H]=-1.9\pm0.25$, in
agreement with the spectroscopic value by Shetrone \etal (2001). No
metallicity gradient is detected along the galaxy.

\acknowledgements

We are indebted to the anonymous referee of this paper for several comments
and, in particular, for suggesting a further discussion of the blue-stars
problem. That point has been largely improved thanks to enlightening comments
from Drs. C. Gallart, G. Piotto, G. Preston and J. M. Torrelles.

This article is based on observations made with the 2.5 m Isaac Newton
Telescope operated on the island of La Palma by the ING in the Spanish
Observatorio del Roque de Los Muchachos, and with the IAC 80 telescope
operated by IAC in the Observatorio del Teide of the Instituto de Astrof\'\i
sica de Canarias. This research has made use of the NASA/IPAC Extragalactic
Database (NED) which is operated by the Jet Propulsion Laboratory, California
Institute of Technology, under contract with the National Aeronautics and
Space Administration. Javier Alonso-Grac\'{\i}a gratefully thanks to the IAC
for their hospitality in La Laguna. This research has made use of the
Digitized Sky Survey, produced at the Space Telescope Science Institute at
Baltimore under U.S.  grant NAGW-2166. This research has been supported by
the Instituto de Astrof\'\i sica de Canarias (grant P3/94), the DGESIC of the
Kingdom of Spain (grant PI97-1438-C02-01), and the DGUI of the autonomous
government of the Canary Islands (grant PI1999/008).

\newpage

\begin{deluxetable}{ccccc}
\tablenum{1}
\tablewidth{400pt}
\tablecaption{Journal of observations
\label{journal}}
\tablehead{
\colhead{Date} & \colhead{Ursa Minor Field} & \colhead{Time (UT)} & \colhead{Filter} &
\colhead{Exp. time (s)}}
\startdata
98.05.27 & A & 22:39 & $B$ & 900 \nl
98.05.27 & A & 22:57 & $B$ & 900 \nl
98.05.27 & A & 23:15 & $R$ & 900 \nl
98.05.27 & A & 23:33 & $R$ & 900 \nl
98.05.27 & B & 23:52 & $B$ & 900 \nl
98.05.27 & B & 00:11 & $B$ & 900 \nl
98.05.27 & B & 00:29 & $R$ & 900 \nl
98.05.27 & B & 00:48 & $R$ & 900 \nl
98.05.28 & A & 21:28 & $B$ &  30 \nl
98.05.28 & A & 21:32 & $R$ &  20 \nl
98.05.28 & B & 21:51 & $R$ &  20 \nl
98.05.28 & B & 21:54 & $B$ &  30 \nl
98.05.28 & C & 22:10 & $B$ &  30 \nl
98.05.28 & C & 22:15 & $R$ &  20 \nl
98.05.28 & C & 22:46 & $B$ & 900 \nl
98.05.28 & C & 23:04 & $B$ & 900 \nl
98.05.28 & C & 23:22 & $R$ & 900 \nl
98.05.28 & C & 23:40 & $R$ & 900 \nl
99.06.15 & A & 22:59 & $V$ &  10 \nl
99.06.15 & A & 23:02 & $V$ &  10 \nl
99.06.15 & A & 23:06 & $V$ & 600 \nl
99.06.15 & A & 23:20 & $I$ &  10 \nl
99.06.15 & A & 23:23 & $I$ & 600 \nl
99.06.15 & A & 23:37 & $B$ & 120 \nl
99.06.15 & A & 23:43 & $R$ & 120 \nl
01.05.13$^*$ & A & 03:23 & $V$ & 900 \nl
01.05.13$^*$ & A & 03:30 & $I$ & 900 \nl
\enddata
$^*$ taken at the IAC-80 telescope at Teide Observatory.
\end{deluxetable}

\newpage

\newpage

\begin{deluxetable}{lcccccc}
\tablenum{2}
\tablewidth{500pt}
\tablecaption{Clusters and parameters used to estimate the distance.
\label{cgg}}
\tablehead{
\colhead{Cluster} & \colhead{Metallicity} & \colhead{$(m-M)_0^{\rm
gc}$} & \colhead{$V^{\rm gc}_{\rm HB,0}$} & \colhead{$V^{\rm gc}_{\rm HB,[Fe/H]}$} & \colhead{$\Delta V_{\rm HB,0}$} & \colhead{$(m-M)^{\rm UMi}_0$}}
\startdata
NGC 4590 & $-2.00\pm0.03$ & $15.23\pm0.16$ & $15.63\pm0.1 $ & $15.67\pm0.13$ & $4.10\pm0.15$ & $19.33\pm0.2$ \nl
NGC 6341 & $-2.16\pm0.02$ & $14.82\pm0.16$ & $15.14\pm0.1 $ & $15.21\pm0.13$ & $4.56\pm0.15$ & $19.38\pm0.2$ \nl
NGC 6397 & $-1.76\pm0.03$ & $12.24\pm0.1 $ & $12.39\pm0.1 $ & $12.39\pm0.13$ & $7.38\pm0.15$ & $19.62\pm0.2$ \nl
NGC 7078 & $-2.02\pm0.04$ & $15.30\pm0.14$ & $15.62\pm0.05$ & $15.67\pm0.1 $ & $4.10\pm0.12$ & $19.40\pm0.2$ \nl
\enddata
\end{deluxetable}

\newpage

\begin{deluxetable}{lcc}
\tablenum{3}
\tablewidth{400pt}
\tablecaption{Morphological and integrated parameters of Ursa Minor
\label{densi_par}}
\tablehead{
\colhead{Parameter} & \colhead{Value} & \colhead{Reference}}
\startdata
$\alpha_{00}$                            & ${\rm 15^h09^m.2}$       & (3) \nl
$\delta_{00}$                            & $67^\circ12\farcm9$      & (3) \nl
$\mu_{''V'',0}$                          & $25.5\pm 0.3$            & (3) \nl  
$r_{\rm c}$                              & $15\farcm 8\pm 1\farcm 2$& (3) \nl
$d$                                      & $70\pm 4$ kpc            & (1) \nl
$[Fe/H]$                                 & $-1.9\pm 0.11$           & (4) (1) \nl
$V_T$                                    & $10.3\pm0.4$             & (2) \nl
$M_{V}$                                  & $-8.9$                   & (1) \nl
$L_{V}$ ($L_\odot$)                      & $3\times 10^5$           & (1) \nl
$M_{VT}$ ($M_\odot$)                     & $23\times 10^6$          & (1) \nl
$M_{VT}/L_{V}$ ($M_\odot$/$L_\odot$)     & $77$                     & (1) \nl
$M_\star(a<18')$ ($M_\odot$)             & $2.25\times 10^5$        & (1) \nl
$M_\star(a<40')$ ($M_\odot$)             & $3.6\times 10^5$         & (1) \nl
$\kappa=1-(M_\star(r<40')/M_{VT})$       & $0.98$                   & (1) \nl
\enddata

\tablerefs{(1) This paper; (2) Caldwell \etal (1992); (3) Irwin \& Hatzidimitriou (1995); (4) Shetrone \etal (2001)}

\end{deluxetable}

\newpage

\newpage

\newpage

\newpage

\newpage

\newpage

\newpage

\newpage

\newpage

\newpage

\newpage

\begin{references}
\reference{} Aaronson, M., \& Mould, J. 1980, \apj, 240, 804

\reference{} Aparicio, A., \& Gallart, C. 1995, \aj, 110, 2105

\reference{} Aparicio, A., Gallart, C., \& Bertelli, G. 1997, \aj, 114, 680

\reference{} Aparicio, A., Carrera, R., \& Mart\'{\i}nez-Delgado, D. 2001, \aj, 122, 2524

\reference{} Aparicio, A. 2001, in {\it Observed HR diagrams and stellar
evolution: the interplay between observational constraints and theory},
ed. J. Fernandes and T. Lejeune, ASP conference series, in press

\reference{} Azzopardi, M., Lequeux, J., \& Westerlund, B. E. 1986, \aap, 161,
232

\reference{} Baade, W. 1963, Evolution of Stars and Galaxies,
ed. C. Payne-Gaposchkin (Cambridge: MIT Press)

\reference{} Bertelli, G., Bressan, A., Chiosi, C., Fagotto, F., \&
Nasi, E. 1994, \aaps, 106, 275

\reference{} Blitz, L. \& Robishaw, T. 2000, \apj, 541, 675

\reference{} Blumenthal, G. R., Faber, S. M., Primack, J. R. \& Rees,
M. J. 1984, \nat, 311, 511

\reference{} Caldwell, N., Armandroff, T. E., Seitzer, P. \& Da Costa,
G. S. 1992, \aj, 103, 840

\reference{} Cardelli, J. A., Clayton, G. C. \& Mathis, J. S. 1989, \apj,
345, 245

\reference{} Carretta, E., \& Gratton, R. 1997, \aaps, 121, 95 (CG)

\reference{} Cudworth, K. M., Olszewski, E. W. \& Schommer, R. A. 1986, \aj,
92, 766

\reference{} Davidge, T. J. 1994, \aj, 108, 2123

\reference{} Dekel, A. \& Silk, J. 1986, \apj, 303, 39

\reference{} Demarque, P., Zinn, R., Lee, Y. W. \& Yi, S. 2000, \aj, 119, 1398

\reference{} Elston, D., \& Silva, D. R. 1992, \aj, 104, 1360

\reference{} Freedman, W. L. 1992, \aj, 104, 1349

\reference{} Frogel, J. A., Blanco, V. M., Cohen, J. G., \& McCarthy,
M. F. 1982, \apj, 252, 133

\reference{} Gallart, C. 2001, in {\it Observed HR diagrams and stellar
evolution: the interplay between observational constraints and theory},
ed. J. Fernandes and T. Lejeune, ASP conference series, in press

\reference{} Gallart, C., Aparicio, A., V\'\i lchez, J. M., 1996, \aj, 112, 1928

\reference{} Gallart, C., Freedman, W. L., Aparicio, A., Bertelli, G., \&
Chiosi, C. 1999, \aj, 118, 2245

\reference{} Hern\'andez, X., Gilmore, G., \& Valls-Gabaud, D. 2000, \mnras,
317, 831

\reference{} Irwin, M. \& Hatzidimitriou, D. 1995, \mnras, 277, 1354

\reference{} Kleyna, J. T., Geller, M. J., Kenyon, S. J., Kurtz, M. J. \&
Thorstensen, J. R. 1998, \aj, 115, 2359 

\reference{} Kroupa, P. Tout, C. A., \& Gilmore, G. 1993, \mnras, 262, 545

\reference{} Landolt, A. U. 1992, \aj, 104, 340

\reference{} Lee, M. G., Freedman, W. L., \& Madore, B. F. 1993 \aj, 106,
964

\reference{} Mart\'\i nez-Delgado, D., \& Aparicio, A. 1997,
\apj, 480, L107

\reference{} Mart\'\i nez-Delgado, D., \& Aparicio, A. 1998, in IAU
Symp. 192, {\sl The stellar content of the Local Group}, ed. Whitelock, \&
Cannon (San Francisco: ASP), 179

\reference{} Mart\'\i nez-Delgado, D., Gallart, C., \& Aparicio, A. 1999a,
\aj, 118, 862

\reference{} Mart\'\i nez-Delgado, D., Aparicio, A. \& Gallart, C. 1999b,
\aj, 118, 2229

\reference{} Mart\'\i nez-Delgado, D., Alonso-Garc\'{\i}a, J., Aparicio, A., \&
G\'omez-Flechoso, M. A. 2001, ApJ, 549, L63

\reference{} Mart\'\i nez-Delgado, D., G\'omez-Flechoso, M. A., Aparicio, A.,
\& Alonso-Garc\'{\i}a, J., 2002, in preparation

\reference{} Mighell, K. J. 1997, \aj, 114, 1458

\reference{} Mighell, K. J. \& Burke, C. J. 1999, \aj, 118, 366

\reference{} Mould, J. R., Cannon, R. D., Frogel, J. A., \& Aaronson, M. 1982, \apj, 254, 500

\reference{} Nemec, J. M., Wehlau, A. \& Mendes de Oliveira, C. 1988, \aj,
96, 528

\reference{} Olsen, K. A. G. 1999, \aj, 117, 2244

\reference{} Olszewski, E. W. \& Aaronson, M. 1985, \aj, 90, 2221

\reference{} Piotto, G., De Angeli, F,. Bono, G,. Cassisi, S., King, I. R.,
Djorgovski, G., \& Meylan, G. 2002, in preparation

\reference{} Preston, G. W. \& Sneden, C. 2000, \aj, 120, 1014

\reference{} Reid, I. N. 1999, \araa, 37, 191

\reference{} Rosenberg, A., Saviane, I., Piotto, G., \& Aparicio, A. 1999,
\aj, 118, 2306

\reference{} Rosenberg, A., Piotto, G., Saviane, I. \& Aparicio, A. 2000a,
\aaps, 144, 5

\reference{} Rosenberg, A., Aparicio, A., Saviane, I. \& Piotto, G. 2000b,
\aaps, 145, 451

\reference{} Saviane, I., Rosenberg, A., Piotto, G., \& Aparicio, A. 2000,
\aap, 355, 966

\reference{} Schlegel, D. J., Finkbeiner, D. P. \& Davis, M. 1998, \apj, 500, 525

\reference{} Shetrone, M. D., C\^ot\'e, P. \& Sargent, W. L. W. 2001,\apj,
548, 592

\reference{} Shu, F. H., Adams, F. C., \& Lizano, S. 1987, \araa, 25, 23

\reference{} Stetson, P. B. \& Harris, W. E. 1988, \aj, 96, 909

\reference{} Stetson, P. B. 1994, \pasp, 106, 250

\reference{} Valls-Gabaud, D., Hern\'andez, X., \& Gilmore, G. 2000. In {\it
Massive Stellar Clusters}. ASP Conference Series, Vol. 211,
p. 270. Ed. A. Lan\c{c}on \& C. M. Boily. ASP 211, 270

\reference{} van der Bergh, S.\& Tammann, G. A. 1991, \araa, 29, 363 

\reference{} Vassiliadis, E. \& Wood, P. R. 1993, \apj, 413, 641 

\reference{} White, S. D. M., \& Rees, M. J. 1978, \mnras, 183, 341

\reference{} Wilson, A. G. 1955, \pasp, 67, 27

\reference{} Young, L. M. 1999, \aj, 117, 1758

\reference{} Young, L. M. 2000, \aj, 119, 188

\reference{} Zinn, R., \& West, M. J. 1984, \apjs, 55, 45 (ZW)

\end{references}
\end{document}